\documentclass[
    ,final            
  ]
  {aipproc}

\layoutstyle{6x9}

\def\cm3{~{\rm cm^{-3}}}

\begin{document}

\title{Fermi acceleration at supernova remnant shocks}


\author{D. Caprioli}{
  address={Department of Astrophysical Sciences, Princeton University, 4 Ivy Ln., Princeton, NJ 08544}
}

\begin{abstract}
 We investigate the physics of particle acceleration at non-relativistic shocks exploiting two different and complementary approaches, namely a semi-analytic modeling of cosmic-ray modified shocks and large hybrid (kinetic protons/fluid electrons) simulations. The former technique allows us to extract some information from the multi-wavelength observations of supernova remnants, especially in the $\gamma$-ray band, while the latter returns fundamental insights into the details of particle injection and magnetic field amplification via plasma instabilities. 
In particular, we present the results of large hybrid simulations of non-relativistic shocks, discussing the properties of the transition from the thermal to the non-thermal component, the spectrum of which turns out to be the power-law predicted by first-order Fermi acceleration. Along with a rather effective magnetic field amplification, we find that more than 20\% of the bulk energy is converted in non-thermal particles, altering significantly the dynamics of the shock and leading to the formation of a precursor. 
\end{abstract}

\maketitle


\section{Introduction}
After many decades of speculation about the possible origin of the Galctic cosmic rays (CRs), recent $\gamma$-ray observations finally provided a direct evidence of the acceleration of hadrons at the forward shock of supernova remnants (SNRs).
In fact, only the $\gamma$-rays from the decay of the neutral pions produced in collisions between relativistic nuclei and the background medium can provide a clear-cut probe of hadron acceleration \citep{DAV94}.

The Fermi satellite, in the GeV band, and the Cherenkov telescopes (HESS, MAGIC, VERITAS, etc.), in the TeV band, are providing us with an unprecedented wealth of information, allowing us to point out two important facts:
\begin{itemize}
\item GeV and TeV spectra are typically as steep as, or steeper than, $E^{-2}$, therefore implying that $\gamma$-rays are likely produced via $\pi_0$ decay (hadronic scenario) and not via inverse Compton scattering of relativistic electrons (leptonic scenario) \citep{gamma}. 
This also forces us to rethink the standard theory for non-linear diffusive shock acceleration (NLDSA) \citep[see, e.g.,][for a review]{malkov-drury01} in order to reconcile theoretical predictions and observations in terms of expected CR spectra \citep{efficiency};   
\item both the morphology and the broadband emission of Tycho show that $\sim10\%$ of the shock kinetic energy is converted into accelerated nuclei with energies up to about 1 PeV \citep{tycho}, thereby confirming that CR acceleration may be efficient in SNRs. 
\end{itemize} 

The mechanism that allows particles to gain energy while being scattered back and forth around a shock was outlined by several authors in the late '70s, when also the first quantitative calculations of particle acceleration were put forward.
If acceleration is as efficient as required to account for the spectrum of Galactic CRs --- 10-30\% of the total SN kinetic energy budget--- the shock dynamics must be strongly affected by the back-reaction of the non-thermal particles.
Because of the pressure in CRs, a precursor where the incoming fluid is slowed down develops upstream of the shock. 
At the same time, the temperature of the downstream fluid must be reduced when CR acceleration is efficient, in that the energy converted into accelerated particles is not available for heating. 

Moreover, the super-Alfv\'enical streaming of CRs ahead of the shock is expected to excite plasma instabilities \citep{bell78a} that may radically change the electromagnetic configuration of the shock  \cite{jumpkin}.
The synchrotron emission detected in young remnants has provided a convincing evidence that at SNR forward shocks the magnetic field is amplified by factor 10-100 with respect to the average Galactic value \citep[e.g.,][]{P+06};
such a \emph{magnetic field amplification} has a crucial back-reaction on the shock structure and on the spectrum of the accelerated particles as well, as put forward in \cite{efficiency}.

The semi-analytical apparatus developed in \cite{efficiency} and references therein proved itself able to account for all the distinctive feature of any other approach to NLDSA \citep{comparison}, therefore providing a proper --- and computationally quick --- tool for investigating acceleration at non-relativistic shocks on the scales relevant for SNR phenomenology.

It is possible to show that, under physically-motivated assumptions about the nature of the excited waves, the net effect of the self-amplified magnetic fields is to make the CR spectrum steeper, eventually reversing the standard prediction that correlates efficient acceleration with spectra flatter than $E^{-2}$  \citep{efficiency}. 
This scenario is particularly appealing, in that it would reconcile $\gamma$-ray observations of SNRs with the theory of NLDSA, outlining a self-consistent picture in which efficient CR acceleration correlates with high levels of magnetization and thereby with rather steep spectra.

Nevertheless, if CR-modified shocks are so sensitive to the development of non-linear plasma instabilities that even the CR spectrum depends on the scattering process,  a heuristic account for the crucial particle-wave interplay and for the particle injection cannot represent a completely satisfactory description of the whole system.  

\section{Hybrid simulations of collisionless shocks}
With the recent $\gamma$-ray observations of SNRs, we have finally entered the era of the quantitative analysis of particle acceleration at shocks; therefore it becomes crucial to understand the processes  that play a key role in the problem, as the ones outlined above.
Such an analysis, in principle, involves all the scales from the Larmor radius of the thermal particles to the diffusion length of the highest-energy particles, which is comparable with the size of the SNR itself.
No single approach can cover this wide range, therefore it becomes of primary importance to couple standard NLDSA techniques with a formalism able to account for the microphysics, possibly from first-principles. 

Particle-in-cell methods \citep[e.g.][]{spitkovsky08} are actually the only viable tool to study the electromagnetic interactions between particles and fields in collisionless shocks and  therefore they are able to provide a self-consistent description of the shock transition, of the particle scattering and of the magnetic field amplification.

However, to mitigate the high computational cost of such simulations, we adopt the so-called \emph{hybrid approach}, which describes protons kinetically and electrons as fluid \citep[see, e.g.,][and references therein]{lipatov02, gs12}.
The hybrid limit is particularly useful for investigating collisionless shocks on intermediate scales without giving up an \emph{a priori} treatment of the phenomena driven by ions.

\begin{figure}
  \includegraphics[height=.3\textheight]{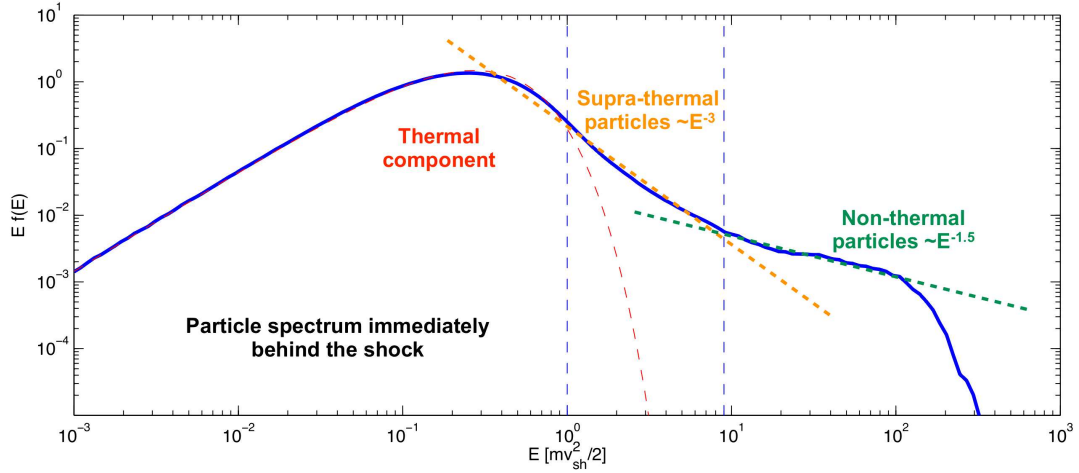}
  \caption{\label{fig} Post-shock ion energy distribution at $t=320\omega_c^{-1}$ for a parallel shock with M=30. 
  The spectrum is integrated in a region of about $200c/\omega_p$ immediately behind the shock.
  The vertical dashed lines mark the boundaries separating thermal, supra-thermal and non-thermal particles, as in the legend. The fraction of the total energy in each of the three components is about 0.75, 0.15 and 0.10, respectively.}
\end{figure}

We perform 2D, non-relativistic simulations by using \emph{dHybrid}, a massively parallel computational code discussed in ref.~\citep{gargate+07}. 
Our setup corresponds to a parallel (initial magnetic field aligned with the shock normal) shock, with both sonic and Alfv\'enic Mach numbers $M=30$ as measured in the downstream (simulation) reference frame. 
We use a box of 200000x1000 cells, corresponding to 40000x200$(c/\omega_p)^2$, a time step of 0.001$\omega_c^{-1}$, with $\omega_p (\omega_c)$ the ion plasma (cyclotron) frequency, and 4 particles per cell.
The shock is generated self-consistently by imposing a perfectly reflecting wall that produces a flow of counter-streaming particles. 
Periodic boundary conditions are set in the transverse direction.
Our main findings are summarized in the following.
\begin{itemize}
\item The magnetic field ahead of the shock is amplified by the streaming of the  accelerated particles, which tend to propagate upstream perturbing the incoming fluid. 
As shown in ref.~\cite{gs12}, the transverse component of the magnetic field may become a few times larger than the parallel initial one, thereby appreciably changing the local field inclination and the effective Alfv\'enic Mach number of the shock.  
\item As time goes by, more and more ions are accelerated to larger and larger energies via multiple interactions with the shock. 
The pressure and energy in these energetic particles lead to the formation of a precursor ahead of the shock and to a progressive decrease of the downstream temperature, $T_{d}$. 
For instance, at $t>300\omega_c^{-1}$, the upstream fluid velocity $V_{sh}$ and $T_d$ are respectively reduced by $\sim10\%$ and $\sim25\%$, with respect to their values at the beginning of the simulation.

\item The ion energy spectrum behind the shock shows three distinct regions (fig.~\ref{fig}):
\begin{itemize}
\item the \emph{thermal} peak, corresponding to a Maxwellian with $T_d\simeq 0.75 T_{sh}$, where  $T_{sh}=\frac{3}{8} E_{sh}/k_B$ is the test-particle prediction and $E_{sh}=\frac{1}{2}m V_{sh}^2$;
\item a \emph{supra-thermal} transition region, with a power-law $\propto E^{-3}$ spanning about one decade in energy above $E_{sh}$ and accommodating $\sim15\%$ of the total energy;
\item a \emph{non-thermal} power-law $\propto E^{-1.5}$, roughly exponentially cut-off around a few hundreds $E_{sh}$ and accommodating  $\sim10\%$ of the total energy. 
\end{itemize}
\end{itemize}

It is very important to stress that the spectrum of the non-thermal component is in perfect agreement with the prediction of the DSA standard theory.
The spectrum in the momentum space, for a strong shock with compression ratio $r=4$, reads $f(p)\propto p^{-3r/(r-1)}=p^{-4}$ therefore, since $4\pi p^2 f(p)dp=f(E)dE$, in the non-relativistic ($E=p^2/2m$) regime of our simulation the spectrum is predicted to be $f(E)\propto E^{-1.5}$.

To our knowledge, the results sketched here represent the first report of such distinctive spectral regions in hybrid simulations of large Mach number collisionless shocks.

Very interestingly, the supra-thermal region contains information about the properties of the particles that have been injected into the acceleration process, and it will be discussed in greater detail in a forthcoming paper. 
The non-thermal part of the spectrum, on the other hand, can be regarded as a probing test of the effectivity of Fermi acceleration at shocks. 
Finally, we stress that here a key role is played here by the extent of the computational box along the shock velocity: smaller boxes, as those used in previous simulations, do not permit the confinement of very energetic particles and the development of the non-thermal tail with the features outlined above.


\begin{theacknowledgments}
The author wishes to thank A. Spitkovsky for his collaboration and L. Gargat\'e for providing a version of \emph{dHybrid}. This research was supported by NSF grant AST-0807381 and NASA grants NNX09AT95G and NNX10A039G. Simulations were performed on the computational resources supported by the PICSciE-OIT TIGRESS High Performance Computing Center and Visualization Laboratory. 
\end{theacknowledgments}

\bibliographystyle{aipproc}   
\bibliography{Gsymp}

\end{document}